\documentclass[preprint, 12pt]{elsarticle}
\usepackage{hyperref}

\hypersetup{colorlinks=true,linkcolor=blue,citecolor=blue,filecolor=blue,urlcolor=blue,}
\usepackage{amsmath}
\usepackage{amssymb}
\usepackage[ruled]{algorithm2e}
\newcommand{\RN}[1]{%
  \textup{\uppercase\expandafter{\romannumeral#1}}%
}
\usepackage{mathtools}

\DeclarePairedDelimiter\floor{\lfloor}{\rfloor}

\journal{Journal of Computational Physics}

\begin{document}

\begin{frontmatter}

\title{High-Order Multiderivative IMEX Schemes}

\author{Alexander J. Dittmann\fnref{myfootnote}\corref{mycorrespondingauthor}}
\address{Department of Astronomy, University of Maryland, College Park, MD 20742-2421}
\cortext[mycorrespondingauthor]{Corresponding author}
\ead{dittmann@astro.umd.edu}

\begin{abstract}
Recently, a 4th-order asymptotic preserving multiderivative implicit-explicit (IMEX) scheme was developed \cite{2020arXiv200108268S}. This scheme is based on a 4th-order Hermite interpolation in time, and uses an approach based on operator splitting that converges to the underlying quadrature if iterated sufficiently. Hermite schemes have been used in astrophysics for decades, particularly for N-body calculations, but not in a form suitable for solving stiff equations. In this work, we extend the scheme presented in \cite{2020arXiv200108268S} to higher orders. Such high-order schemes offer advantages when one aims to find high-precision solutions to systems of differential equations containing stiff terms, which occur throughout the physical sciences. We begin by deriving Hermite schemes of arbitrary order and discussing the stability of these  formulas. Afterwards, we demonstrate how the method of \cite{2020arXiv200108268S} generalises in a straightforward manner to any of these schemes, and prove convergence properties of the resulting IMEX schemes. We then present results for methods ranging from 6th to 12th order and explore a selection of test problems, including both linear and nonlinear ordinary differential equations and Burgers' equation. To our knowledge this is also the first time that Hermite time-stepping methods have been applied to partial differential equations. We then discuss some benefits of these schemes, such as their potential for parallelism and low memory usage, as well as limitations and potential drawbacks.
\end{abstract}
\end{frontmatter}

\section{Introduction}
Numerous physical systems occur in nature that are difficult to model numerically because of stiff terms in the differential equations that govern their evolution. For example, when simulating the evolution of fluid flows that include chemical reactions, cooling, or viscosity, the time step required for stability can be reduced by orders of magnitude when using explicit methods alone \cite[e.g.][]{{1990JCoPh..86..187L},{LANGTANGEN20021125},{2008A&A...488..429T}}. Stiff differential equations also appear in combustion models \cite{book:667288}, and circuitry \cite{1927Natur.120..363V}, for example. If one wishes to efficiently carry out high-accuracy simulations of such systems, especially in multiple dimensions, high-order numerical methods are required. This work explores high-order schemes constructed using Hermite interpolation in time, adapting them to systems where one may use operator splitting to separate stiff and non-stiff terms. 

Hermite schemes were introduced in celestial mechanics as an alternative to Adams-Bashforth-Moulton predictor-corrector schemes \cite{1991ApJ...369..200M} and can be thought of as a multistep generalisation of earlier schemes based on Taylor series \cite{1974nsod.conf..451L}. As the name implies, Hermite schemes are based on constructing a Hermite interpolation polynomial in time. Thus, it follows that by calculating up to the $(q-1)th$-order derivative of a function $f$ at $r$ points in time, one can construct a Hermite interpolation polynomial of order $(qr-1)$\cite{1991ApJ...369..200M}. Typically two points in time are used, trading non-vectorizable operations involving multiple time steps for additional vectorizable operations calculating derivatives of $f$ when compared to Adams-Bashforth-Moulton schemes. 

Various high-order time-stepping algorithms exist for solving differential equations. Runge-Kutta methods of orders higher than four typically involve either very complex Butcher tableaus or tens of stages \cite{2001SIAMR..43...89G,doi:10.1137/07070485X}. Another family of schemes, based on deferred corrections in a Picard integral formalism, requires large numbers of stages and multiple iterations to reach high orders \cite{Dutt2000, christlieb2009}. Thus, high-order algorithms using values only at the beginning and end of an interval offer potential improvements. 

The schemes in this work are largely based on previous investigations focused on N-body algorithms \cite{2016RIKEN.K.N,10.1093/mnras/staa1631,2008NewA...13..498N}, and readers are encouraged to reference those works for further details concerning deriving the schemes presented herein. These schemes are usually applied using an explicit Taylor series predictor, utilising derivatives of the interpolation polynomial at the end of the previous step, followed by a correction step (see, for example, the appendix of \cite{2008NewA...13..498N}). In general, this style of application is not suitable for solving stiff equations. However, Hermite quadrature rules can also be applied in a fully implicit manner, as well as the operator-split manner described in \cite{2020arXiv200108268S}. To elucidate this, we follow \cite{2020arXiv200108268S} and consider the coupled set of differential equations
\begin{equation}\label{test prob}
y'(t) = z(t),~ z'(t)=\frac{g(y(t),z(t))}{\epsilon},~ 0\leq t \leq T,
\end{equation}
where $g:\mathbb{R} ^2 \to \mathbb{R}$ is a smooth function and we choose $0<\epsilon < 1$. These equations are supplemented by initial conditions at $t=0$
\begin{equation}
y(0)=y_0,~ z(0)=z_0.
\end{equation}
The stiff relaxation parameter $\epsilon$ causes Equation (\ref{test prob}) to be a singularly perturbed equation, which exhibits multiscale structure that is challenging for many explicit methods. These problems are well documented in the literature, and a number of books discuss the analysis and application of these equations \cite{{book:667288},{book:266628},{book:1123732}}.

To develop more general schemes, we follow \cite{2020arXiv200108268S} and also consider the following generic system of differential equations with an additive right-hand side
\begin{equation}\label{testprob2}
\frac{du}{dt} = \phi(u) = \phi_E(u) + \phi_I(u),
\end{equation}
where $\phi_E$ and $\phi_I$ contain the non-stiff and stiff terms respectively. 

When problems can be split in the manner of Equation (\ref{testprob2}), the stiff components may be treated implicitly and non-stiff parts explicitly, greatly simplifying the implicit part of the calculation. This procedure leads to implicit-explicit (IMEX) methods. Popular families include those based on Runge-Kutta and multistep methods. Recently, \cite{2020arXiv200108268S} introduced a multiderivative IMEX method, which effectively implements a Hermite scheme for stiff problems. Because of its connection to Hermite schemes, it is straightforward to generalise the method presented in \cite{2020arXiv200108268S} to higher orders by drawing on previous advances in N-body methods \cite{2008NewA...13..498N, 10.1093/mnras/staa1631}. Developing these high-order methods is the primary goal of this paper, but before doing so it is necessary to introduce Hermite quadrature schemes of arbitrary order. Afterwards, we show how these can be used to generalise the multiderivative IMEX scheme of \cite{2020arXiv200108268S}. We provide a few theoretical results on their convergence, and apply these Hermite IMEX schemes to example stiff ordinary and partial differential equations.

\section{Hermite Schemes}

Instead of repeating the entirety of derivations found elsewhere, we state a number of results \cite{2016RIKEN.K.N,10.1093/mnras/staa1631} and proceed to discussing stability properties. Let us consider the equation 
\begin{equation}
\frac{du}{dt}=f(u, t)
\end{equation}
and a time step $\Delta t$, defining $h\equiv \Delta t/2$. Then, let us define a fitting polynomial at the centre of the interval $[t-h, t+h]$ using the notation $f^{(n)}\equiv\frac{d^n}{dt^n}f$
\begin{equation}
F_n=\frac{h}{n!}f^{(n)}(u(t), t),
\end{equation}
which will be determined by linear combinations of values at the beginning end of the interval
\begin{equation}
F_n^\pm=\frac{1}{2}\frac{h^n}{n!}\left(f^{(n)}(u(t+h), t+h)\pm f^{(n)}(u(t-h), t-h)\right).
\end{equation}
We can then calculate an update to $u$, $\Delta u \equiv u(t+h)-u(t-h)$, as
\begin{equation}\label{update}
\Delta u = \int_{-h}^{h}f(t)dt\approx
\left(F_0 + \frac{1}{3}F_2+ ... + \frac{1}{2p+1}F_{2p} \right)\Delta t,
\end{equation}
where the series is truncated at $n=2p$ and the order of the scheme is $2(p+1)$. Then, we can solve for $F_n$ in terms of linear combinations of $F_n^\pm$ using
\begin{equation}
\begin{bmatrix}
\textbf{F}^+_0 \\
\textbf{F}^-_1 \\
\textbf{F}^+_2 \\
\textbf{F}^-_3 \\
\vdots \\
\end{bmatrix}
=
\begin{bmatrix}
\left(\begin{smallmatrix} 0\\0 \end{smallmatrix}\right) &\left(\begin{smallmatrix} 2\\0 \end{smallmatrix}\right) &
\left(\begin{smallmatrix} 4\\0 \end{smallmatrix}\right) &\left(\begin{smallmatrix} 6\\0 \end{smallmatrix}\right) & \cdots \\
0 &\left(\begin{smallmatrix} 2\\1 \end{smallmatrix}\right) &
\left(\begin{smallmatrix} 4\\1 \end{smallmatrix}\right) &\left(\begin{smallmatrix} 6\\1 \end{smallmatrix}\right) & \cdots \\
0 &\left(\begin{smallmatrix} 2\\2 \end{smallmatrix}\right) &
\left(\begin{smallmatrix} 4\\2 \end{smallmatrix}\right) &\left(\begin{smallmatrix} 6\\2 \end{smallmatrix}\right) & \cdots \\
0 & 0 & \left(\begin{smallmatrix} 4\\3 \end{smallmatrix}\right) &\left(\begin{smallmatrix} 6\\3 \end{smallmatrix}\right) & \cdots \\
\vdots & \vdots & \vdots & \vdots & \ddots \\
\end{bmatrix}
\begin{bmatrix}
\textbf{F}_0 \\
\textbf{F}_2 \\
\textbf{F}_4 \\
\textbf{F}_6 \\
\vdots \\
\end{bmatrix}
=\bar{A}
\begin{bmatrix}
\textbf{F}_0 \\
\textbf{F}_2 \\
\textbf{F}_4 \\
\textbf{F}_6 \\
\vdots \\
\end{bmatrix}
\end{equation}
where $\bar{A}$ is a matrix constructed from the even columns of an upper triangular Pascal matrix, and parentheses indicate binomial coefficients. Thus, for a $2(p+1)$th-order scheme with $p+1$ terms, 
\begin{equation}
\Delta u = \left( c^p_0F^+_0 + c^p_1F^-_1 + ...+ c^p_pF^\pm_p \right)\Delta t, 
\end{equation}
where $c^p_k$ is defined for $0\leq k \leq p$ as
\begin{equation}\label{cpk}
c^p_k=\frac{1}{(-2)^k}\frac{(2k)!!}{(2k+1)!!}\left(\begin{smallmatrix} p-k+m\\p-k \end{smallmatrix}\right)\frac{(2k+1)!!}{(2k+1-2m)!!}\frac{(2p+1-2m)!!}{(2p+1)!!},
\end{equation}
where $m\equiv \floor*{(k+1)/2}$\cite{2016RIKEN.K.N}. As an example, let us consider the resulting 4th-order scheme
\begin{equation}\label{4thord}
\Delta u = \frac{\Delta t}{2}\left(f_1 + f_0\right) - \frac{\Delta t^2}{12}\left(f^{(1)}_1 - f^{(1)}_0\right)
\end{equation}
where the subscript 0 indicates a value at the beginning of a time step and the subscript 1 indicates a value at the end of a time step. Explicit forms up to 12th order are provided in \ref{apndx:A}. Typically these schemes are applied in a predictor-corrector fashion using an explicit Taylor series predictor, constructed using derivatives of the interpolating polynomial evaluated at the end or the previous interval.
However, this procedure is not suitable for stiff equations, and we will instead follow \cite{2020arXiv200108268S}, using a mix of lower-order forward and backward Taylor series as a predictor. But first, we consider the stability of the underlying Hermite schemes.

Let us apply Equation (\ref{4thord}) to the test function $\frac{du}{dt}=f(u)=\lambda u, ~ f^{(n)}(u)=\lambda^nu$: 
\begin{equation}
\Delta u = u_1 - u_0 = \frac{\Delta t}{2}\left(\lambda u_1 + \lambda u_0\right) - \frac{\Delta t^2}{12}\left(\lambda^2u_1 - \lambda^2u_0\right).
\end{equation}
Rearranging, and defining $\Phi(\lambda\Delta t)\equiv u_1/u_0$ we see that 
\begin{equation}\label{eqphi}
\Phi(K\Delta t) = \frac{1 + \frac{1}{2}\lambda\Delta t + \frac{1}{12}\lambda^2\Delta t^2   }{1 - \frac{1}{2}\lambda\Delta t + \frac{1}{12}\lambda^2\Delta t^2   }.
\end{equation}
Analogously to the trapezoidal rule, we see that for any $\lambda<0$ and $\Delta t>0$, $|\Phi(\lambda\Delta t)|<1$. Thus, the 4th-order method is A-stable, with a linear stability region consisting of the left half of the complex plane.  
Now we consider $\Phi(\lambda\Delta t)$ for higher-order methods. Based on the sign of the coefficients $c^p_k$ in Equation (\ref{cpk}), we can see that any odd power of $\Delta t$ will be accompanied by a negative coefficient in the denominator of $\Phi(\lambda\Delta t)$ but a positive coefficient in the numerator. Thus, all Hermite schemes are A-stable. 

\subsection{Hermite IMEX schemes}
In this section, we show how Hermite schemes can be used to construct high-order IMEX schemes in the manner of \cite{2020arXiv200108268S}. Concerning notation, we represent the $m$-th time derivative of $g(y(t_n),z(t_n))$ as $g^{(m)}_n$. As an example, let us first consider the 6th-order Hermite scheme as a base for solving Equation (\ref{test prob}),
\begin{equation}
\begin{split}
\Delta y = \frac{\Delta t}{2}(z_1+z_0)-\frac{\Delta t^2}{10}(g_1-g_0)+\frac{\Delta t^3}{120}(g^{(1)}_1+g^{(1)}_0) \\
\Delta z = \frac{\Delta t}{2\epsilon}(g_1+g_0)-\frac{\Delta t^2}{10\epsilon}(g^{(1)}_1-g^{(1)}_0)+\frac{\Delta t^3}{120\epsilon}(g^{(2)}_1+g^{(2)}_0).
\end{split}
\end{equation}

The corresponding Hermite IMEX method is as follows.

\SetAlgoNoLine
\LinesNumberedHidden
\begin{algorithm}[H] \label{alg 1}
\begin{enumerate}
    \item \textbf{Predict.}  Given the solution $(y_n,z_n)$, we compute a 3rd-order IMEX Taylor approximation
    \begin{equation}
    y_{[0]} = y_n + \Delta tz_n + \frac{\Delta t^2}{2\epsilon}g_n + \frac{\Delta t^3}{6\epsilon}g^{(1)}_n
    \end{equation}
    \begin{equation}
    z_{[0]} = z_n + \frac{\Delta t}{\epsilon} g_{[0]} - \frac{\Delta t^2}{2\epsilon}g^{(1)}_{[0]} + \frac{\Delta t^3}{6\epsilon}g^{(2)}_{[0]}
    \end{equation}
    for the unknowns $y_{[0]}$ and $z_{[0]}$ that will be initial guesses for our approximation to $y_{n+1}$ and $z_{n+1}$. For the 6th-order Hermite IMEX scheme, we use a 3rd-order forward Taylor series in $y$ and a 3rd-order backwards Taylor series in $z$. For a $(2n)$th-order Hermite IMEX scheme we would use an $n$th-order Taylor series in this step.
    \item \textbf{Correct.} Based on the initial step, for $0\leq k \leq k_{\rm max}-1$ we solve
    \begin{equation}
    y_{[k+1]} = y_n + \frac{\Delta t}{2}(z_{[k]} + z_n) - \frac{\Delta t^2}{10\epsilon}(g_{[k]}-g_n) + \frac{\Delta t^3}{120\epsilon}(g^{(1)}_{[k]} + g^{(1)}_n)
    \end{equation}
    \begin{equation}
    \begin{split}
    z_{[k+1]} = z_n + \frac{\Delta t}{2\epsilon}(g_{[k]} + g_n) - \frac{\Delta t^2}{10\epsilon}(g^{(1)}_{[k]}-g^{(1)}_n) + \frac{\Delta t^3}{120\epsilon}(g^{(2)}_{[k]} + g^{(2)}_n) \\
    +\frac{\Delta t}{\epsilon}(g_{[k+1]} - g_{[k]}) - \frac{\Delta t^2}{2\epsilon}(g^{(1)}_{[k+1]} - g^{(1)}_{[k]})
    + \frac{\Delta t^3}{6\epsilon}(g^{(2)}_{[k+1]} - g^{(2)}_{[k]} )
    \end{split}
    \end{equation}
    for $y_{[k_{\rm max}]}$ and $z_{[k_{\rm max}]}$. 
    \item \textbf{Update.} The update for the solution is then carried out as \\$y_{n+1}=y_{[k_{\rm max}]}$,~  $z_{n+1}=z_{[k_{\rm max}]}$.
\end{enumerate}
\caption{For the solution of Equation (\ref{test prob}), we propose the following 6th-order Hermite IMEX method to advance the solution from $t_n$ to $t_{n+1}$}.
\end{algorithm}
Note that the corrector for $y$ is simply the Hermite corrector, and that as $y_{[k]}$ and $z_{[k]}$ converge, $g^{(m)}_{[k+1]}=g^{(m)}_{[k]}$ and the corrector for $z$ converges to the Hermite scheme. Thus, it is simple to implement any of the methods in \ref{apndx:A} or the natural extensions thereof in IMEX form. Subsequently, we will extend this algorithm to arbitrary splittings, following \cite{2020arXiv200108268S}. We shall do so for the 8th-order Hermite IMEX scheme in order to demonstrate how easily one can translate a given Hermite scheme to a Hermite IMEX scheme. 

One may note that higher-order derivatives of $g$ are used to calculate the update to $z$ than for $y$, and that one has enough information to use a higher-order quadrature when updating $y$, although this would not change the overall order of the scheme. However, when modelling physical systems, there are often physical quantities that ought to be conserved, but are not conserved by a given algorithm. One such system is a Keplerian orbit, where the Laplace-Runge-Lenz vector should be conserved, but is not when the system is treated by many time-stepping algorithms such as the one outlined in Algorithm \ref{alg 1}, even when those algorithms conserve energy and the scalar eccentricity of the orbit. In such cases one may tune the truncation error of $y$ using higher-order derivatives of $g$ in order to improve the conservation of such quantities, as outlined in \cite{10.1093/mnras/staa1631}.

We now consider an 8th-order Hermite IMEX scheme, using Equation (\ref{testprob2}). 
\SetAlgoNoLine
\LinesNumberedHidden
\begin{algorithm}[H] \label{alg 2}
\begin{enumerate}
    \item \textbf{Predict.}  Given the solution $u_n$, we compute a 4th-order IMEX Taylor approximation to $u_{[0]}$,
    \begin{equation}
    \begin{split}
    u_{[0]} = u_n + \Delta t \left(\phi_E(u_n) + \phi_I(u_{[0]})\right) + \frac{\Delta t^2}{2}\left(\phi_E^{(1)}(u_n)-\phi_I^{(1)}(u_{[0]})\right) \\+ \frac{\Delta t^3}{6}\left(\phi_E^{(2)}(u_n) + \phi_I^{(2)}(u_{[0]})\right) + \frac{\Delta t^4}{24}\left(\phi_E^{(3)}(u_n) - \phi_I^{(3)}(u_{[0]})\right),
    \end{split}
    \end{equation}
    performing a forward Taylor expansion for $\phi_E$ and a backward Taylor expansion for $\phi_I$.
    \item \textbf{Correct.} Based on the initial step, for $0\leq k \leq k_{\rm max}-1$ we solve
    \begin{equation}
    \begin{split}
    u_{[k+1]} = u_n + \Delta t\left(\phi_I(u_{[k+1]})-\phi_I(u_{[k]})\right)\\- \frac{\Delta t^2}{2}\left(\phi_I^{(1)}(u_{[k+1]}) - \phi_I^{(1)}(u_{[k]})\right)
    + \frac{\Delta t^3}{6}\left(\phi_I^{(2)}(u_{[k+1]}) - \phi_I^{(2)}(u_{[k]})\right)\\
    - \frac{\Delta t^4}{24}\left( \phi_I^{(3)}(u_{[k+1]})-\phi_I^{(3)}(u_{[k]})\right)+\frac{\Delta t}{2} \left(\phi(u_{[k]}+\phi(u_n)\right)\\ - \frac{3 \Delta t^2}{28}\left(\phi^{(1)}(u_{[k]}) -\phi^{(1)}(u_n)\right) + \frac{\Delta t^3}{84}\left(\phi^{(2)}(u_{[k]}) + \phi^{(2)}(u_n)\right)\\ - \frac{\Delta t^4}{1680}\left(\phi^{(3)}(u_{[k]} - \phi^{(3)}(u_n) \right)
    \end{split}
    \end{equation}
    for $u_{[k_{\rm max}]}$.
    \item \textbf{Update.} The update for the solution is then defined as \\$u_{n+1}=u_{[k_{\rm max}]}$.
\end{enumerate}
\caption{For general problems, we propose the following 8th order Hermite IMEX method to advance the solution of Equation (\ref{testprob2}) from $t_n$ to $t_{n+1}$, emphasising that one may construct a Hermite IMEX scheme of any even order in this manner. } 
\end{algorithm}

Note that in order to construct this scheme, one simply includes one more term in the Taylor series during prediction and correction, and uses a higher-order Hermite scheme during the correction step.
We emphasise that, similarly to the 4th-order method presented in \cite{2020arXiv200108268S}, intermediate steps do not need to be stored, and that the algorithm only requires storage of values at $t_n$ and $t_{n+1}$. This is advantageous over multistep methods as well as most Runge-Kutta methods. 

\subsection{Accuracy each iteration}
In this section, we review how the accuracy of the approximate solution $z_{[k_{\rm max}]}$ improves each iteration. It is clear that Algorithms \ref{alg 1} and \ref{alg 2} have the same order of accuracy as their predictor when $k_{\rm max}=0$, and approach the order of accuracy of the underlying Hermite quadrature rule as $k_{\rm max} \rightarrow \infty$. Here, we show that with each iteration $k>0$ the approximate solution gains one order of accuracy, up to the order of the underlying Hermite quadrature. This result has been demonstrated for the 4th-order scheme \cite{2020arXiv200108268S}, but we repeat this exercise here for the 6th-order scheme, and comment on how the same result holds for implementations of any order. 

First, we lay down a few preliminaries. As an example, we take the update to $z$ as in Algorithm \ref{alg 1}, although the same result holds for $y$ and for problems cast in the form of Algorithm \ref{alg 2}. First, we assume that $g(y,z)$ and its derivatives are Lipschitz continuous, such that 
\begin{equation}
\begin{split}
\|g(y_1,z_1)-g(y_2,z_2)\|\leq L_g \|y_1-y_2, z_1-z_2\|,\\\|g^{(1)}(y_1,z_1)-g^{(1)}(y_2,z_2)\|\leq L_{g^{(1)}}\|y_1-y_2, z_1-z_2\|,\\\|g^{(2)}(y_1,z_1)-g^{(2)}(y_2,z_2)\|\leq L_{g^{(2)}}\|y_1-y_2, z_1-z_2\|,
\end{split}
\end{equation}
where $L_g, L_{g^{(1)}},$ and $ L_{g^{(2)}}$ are constants that subsume the respective Lipschitz constants and factors of $\epsilon$, and Euclidean norms are denoted by $\|x\|$.

Then, consider a 6th-order Hermite quadrature rule for a generic function $f$, such that
\begin{equation}\label{quad6}
\mathcal{I}[f_n,f_{[k]}]=\frac{\Delta t}{2}\left(f_{[k]}+f_n \right) - \frac{\Delta t^2}{10}\left(f^{(1)}_{[k]}-f^{(1)}_n \right) + \frac{\Delta t^3}{120}\left(f^{(2)}_{[k]}+f^{(2)}_n \right)
\end{equation}
and such that
\begin{equation}
\mathcal{I}[g_n,g_{n+1}]=\int_{t_n}^{t_{n+1}}g(y,z)dt + \mathcal{O}(\Delta t^7).
\end{equation}
In this notation, one updates $z$ according to 
\begin{equation}
\begin{split}
z_{[k+1]}=z_n + \frac{\Delta t}{\epsilon}\left(g_{[k+1]}-g_{[k]}\right) - \frac{\Delta t^2}{2\epsilon}\left(g^{(1)}_{[k+1]}-g^{(1)}_{[k]}\right)\\ + \frac{\Delta t^3}{6\epsilon}\left(g^{(2)}_{[k+1]}-g^{(2)}_{[k]}\right) + \frac{1}{\epsilon}\mathcal{I}[g_n,g_{[k]}].
\end{split}
\end{equation}
Then, assuming that $z_{n+1}$ is the true solution at time $t_{n+1}$, the error of the approximate solution $z_{[k]}$ is $\delta_{z:k}=z_{[k]}-z_{n+1}$ and the error of $y_{[k]}$ is $\delta_{y:k}=y_{[k]}-y_{n+1}$. Then, the error of the approximate solution is $\delta_k=\|(\delta_{y:k}, \delta_{z:k})\|$. Thus, the error of the quadrature rule given by Equation (\ref{quad6}) at iteration $k$ is given by 
\begin{equation}
\begin{split}
\left|\mathcal{I}[z_n,z_{n+1}]-\mathcal{I}[z_n,z_{[k]}]\right| \\ =\left|\frac{\Delta t}{2}\left(z_{n+1} - z_{[k]} \right)-\frac{\Delta t^2}{10}\left(z^{(1)}_{n+1} - z^{(1)}_{[k]}\right) + \frac{\Delta t^3}{120}\left(z^{(2)}_{n+1} - z^{(2)}_{[k]}\right) \right|\\
\leq \frac{\Delta t}{2}\left|z_{n+1} - z_{[k]} \right|+\frac{\Delta t^2}{10}\left|z^{(1)}_{n+1} - z^{(1)}_{[k]}\right| + \frac{\Delta t^3}{120}\left|z^{(2)}_{n+1} - z^{(2)}_{[k]}\right|\\
\leq \frac{\Delta t}{2}\delta_k + \frac{\Delta t^2}{10}L_g\delta_k+\frac{\Delta t^3}{120}L_{g^{(1)}}\delta_k
\end{split}
\end{equation}
and
\begin{equation}
\begin{split}
\left|\mathcal{I}[g_n,g_{n+1}]-\mathcal{I}[g_n,g_{[k]}]\right|\leq \frac{\Delta t}{2}L_g\delta_k + \frac{\Delta t^2}{10}L_{g^{[1]}}\delta_k+\frac{\Delta t^3}{120}L_{g^{(2)}}\delta_k
\end{split}
\end{equation}
for integrating $z$ and $g$ respectively. Then, since the true solution satisfies 
\begin{equation}
z_{n+1} = z_n + \frac{1}{\epsilon}\int_{t_n}^{t_{n+1}}g(y,z)dt,
\end{equation}
the error of the approximate solution is given by 
\begin{equation}
\begin{split}
|\delta_{z:k+1}|=\left|\frac{\Delta t}{\epsilon}\left(g_{[k+1]}-g_{[k]}\right)+\frac{\Delta t^2}{2\epsilon}\left(g^{(1)}_{[k+1]}-g^{(1)}_{[k]}\right)\right.\\ \left.
+\frac{\Delta t^3}{6\epsilon}\left(g^{(2)}_{[k+1]}-g^{(2)}_{[k]}\right)
+\frac{1}{\epsilon}\mathcal{I}[g_n,g_{n+1}] - \frac{1}{\epsilon}\int_{t_n}^{t_{n+1}}g(y,z)dt \right|\\
\leq \frac{\Delta t}{\epsilon}\underbrace{\left|g_{[k+1]}-g_{[k]}\right|}_\text{\RN{1}} + \frac{\Delta t^2}{2\epsilon}\underbrace{\left|g^{(1)}_{[k+1]}-g^{(1)}_{[k]}\right|}_\text{\RN{2}} + \frac{\Delta t^3}{6\epsilon}\underbrace{\left|g^{(2)}_{[k+1]}-g^{(2)}_{[k]}\right|}_\text{\RN{3}}\\ + \frac{1}{\epsilon}\underbrace{\left|\mathcal{I}[g_n,g_{[k]}] - \int_{t_n}^{t_{n+1}}g~dt
\right|}_\text{\RN{4}}.
\end{split}
\end{equation}
We treat each part separately,
\begin{equation}
\RN{1}\leq |g_{n+1}-g_{[k+1]}| + |g_{n+1} - g_{[k]}| \leq L_g\delta_k + L_g\delta_{k+1},
\end{equation}
\begin{equation}
\RN{2}\leq |g^{(1)}_{n+1}-g^{(1)}_{[k+1]}| + |g^{(1)}_{n+1} - g^{(1)}_{[k]}| \leq L_{g^{(1)}}\delta_k + L_{g^{(1)}}\delta_{k+1},
\end{equation}
\begin{equation}
\RN{3}\leq |g^{(2)}_{n+1}-g^{(2)}_{[k+1]}| + |g^{(2)}_{n+1} - g^{(2)}_{[k]}| \leq L_{g^{(2)}}\delta_k + L_{g^{(2)}}\delta_{k+1},
\end{equation}
\begin{equation}
\begin{split}
\RN{4}\leq \left|\mathcal{I}[g_n,g_{[k]}]-\mathcal{I}[g_n,g_{n+1}]\right| + \left| \mathcal{I}[g_n,g_{n+1}]-\int_{t_n}^{t_{n+1}}g~dt
\right|\\
\leq \frac{\Delta t}{2}L_g\delta_k + \frac{\Delta t^2}{10}L_{g^{(1)}}\delta_k + \frac{\Delta t^3}{120}L_{g^{(2)}}\delta_k + \mathcal{O}(\Delta t^7).
\end{split}
\end{equation}
Thus, 
\begin{equation}
\begin{split}
|\delta_{z:k+1}| \leq \frac{\Delta t}{\epsilon}L_g\left(\delta_k  + \delta_{k+1}  \right) + \frac{\Delta t^2}{2\epsilon}L_{g^{(1)}}\left(\delta_k  + \delta_{k+1}  \right) \\
+  \frac{\Delta t^3}{6\epsilon}L_{g^{(2)}}\left(\delta_k  + \delta_{k+1}  \right)
+  \frac{\Delta t}{2\epsilon}L_{g}\left(\delta_k\right)
+  \frac{\Delta t^2}{10\epsilon}L_{g^{(1)}}\left(\delta_k\right)\\
+  \frac{\Delta t^3}{120\epsilon}L_{g^{(2)}}\left(\delta_k\right) + \mathcal{O}(\Delta t^7).
\end{split}
\end{equation}
Note that regardless of the order of the Hermite quadrature, the limiting term determining the order of $\delta_{z:k}$ each iteration only changes by one factor of $\Delta t$ until it reaches the underlying order of the quadrature rule. That is to say, no matter what order the overall scheme, the order of the algorithm at iteration $k+1$ is given by 
\begin{equation}
|\delta_{k+1}| = \mathcal{O}(\Delta t \delta_{k+1}) + \mathcal{O}(\Delta t \delta_k) + \mathcal{O}(\Delta t^7).
\end{equation}
Thus, for a $(2n)th$-order scheme with an $(n)th$-order accurate predictor, as described above, $k_{\rm max}=n$ is sufficient to achieve the maximal order of the algorithm. However, for some problems where these algorithms experience order reduction, or for problems where the underlying system has symmetry in time that is preferable to maintain numerically, more iterations may be useful \cite{{2020arXiv200108268S},{10.1093/mnras/staa1631}}. 

\section{Numerical Experiments}
In this section, we apply Hermite IMEX schemes of order 6--12 to a selection of test problems. We begin in Section \ref{VanDerPolOsc} by examining the van der Pol oscillator and making comparisons to the 4th-order algorithm with respect to efficiency. We find that higher-order methods are more susceptible to order reduction, particularly for very stiff problems. For very stiff problems, the 4th-order method may be preferable to higher-order algorithms because it appears less susceptible to order reduction. 
In Section \ref{Linear}, we apply Hermite IMEX methods to a stiff linear ODE and find less-significant order reduction. When applying the 6th-order scheme to Burgers' equation in Section \ref{Burgers}, we observer order reduction in cases when the time scale for the diffusive term is more than $\sim 30$ times smaller than that for the convective term.

\subsection{Van der Pol oscillator}\label{VanDerPolOsc}
To directly compare our methods to that presented in \cite{2020arXiv200108268S}, we consider the van der Pol oscillator, which has the form of Equation (\ref{test prob}) with 
\begin{equation}
g(y, z) = (1-y^2)z - y.
\end{equation}
We use the same initial conditions as \cite{2020arXiv200108268S} when testing the 6th-order method, 
\begin{equation}
y_0 = 2, ~ z_0 = -\frac{2}{3} + \frac{10}{81}\epsilon - \frac{292}{2187}\epsilon^2,
\end{equation}
which are a common choice in the literature \cite[e.g.][]{{book:266628},{BOSCARINO20091515}} and facilitate direct comparison with the results of \cite{2020arXiv200108268S}. When testing the 8th-order method, we use the initial conditions 
\begin{equation}
y_0 = 2, ~ z_0 = -\frac{2}{3} + \frac{10}{81}\epsilon - \frac{292}{2187}\epsilon^2 + \frac{15266}{59049}\epsilon^3,
\end{equation}
which ensures that $g^{(3)}(y_0,z_0)\rightarrow0$ as $\epsilon \rightarrow0$, such that the system recovers the correct behavior in the asymptotic limit.
We integrate these equations until $t_{\rm end}=0.5$. Errors are calculated by comparing $y(t_{\rm end})$ and $z(t_{\rm end})$ to reference values $y_{\rm ref}$ and $z_{\rm ref}$, computed using an additional integration using a time step half as small as that for the shortest-timestep result presented. We compare the values at $t_{\rm end}$ to the reference values using the Euclidean norm $\delta\equiv \|y(t_{\rm end}) - y_{\rm ref}, z(t_{\rm end}) - z_{\rm ref}\|$.

\begin{figure}
\includegraphics[width=\columnwidth]{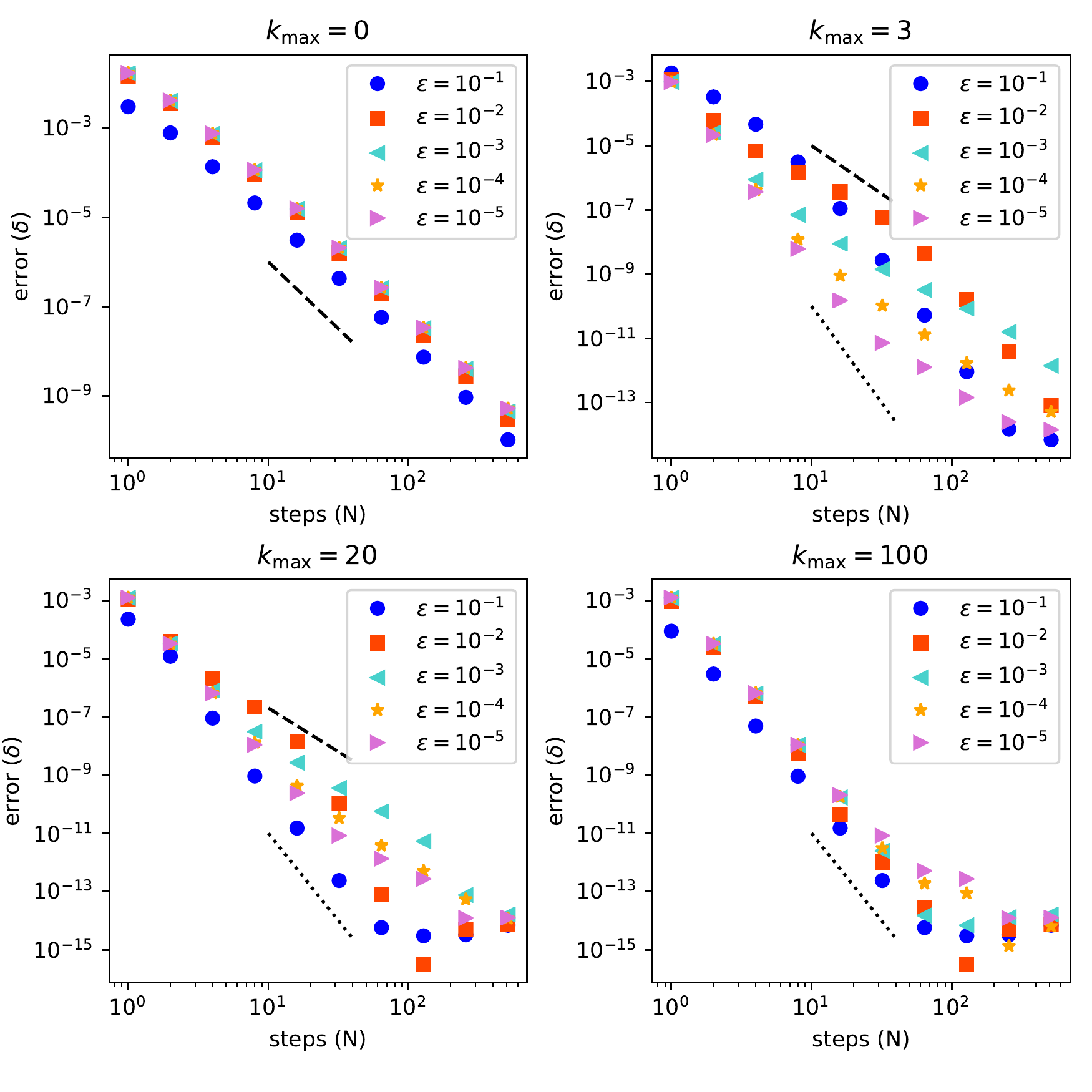}
\caption{The convergence of 6th-order Hermite IMEX schemes applied to the van der Pol equation. Dashed black lines indicate a 3rd-order scaling, and dotted black lines indicate a 6th-order scaling. 
Results for $\epsilon=10^{-1}$ are plotted using blue circles, results for  $\epsilon=10^{-2}$ are plotted using vermilion squares, results for $\epsilon=10^{-3}$ are plotted using turquoise left-pointing triangles, results for $\epsilon=10^{-4}$ are plotted using orange stars, and results for $\epsilon=10^{-5}$ are plotted using purple right-facing triangles.}
\label{fig:HIMEX-6}
\end{figure}

\begin{figure}
\includegraphics[width=\columnwidth]{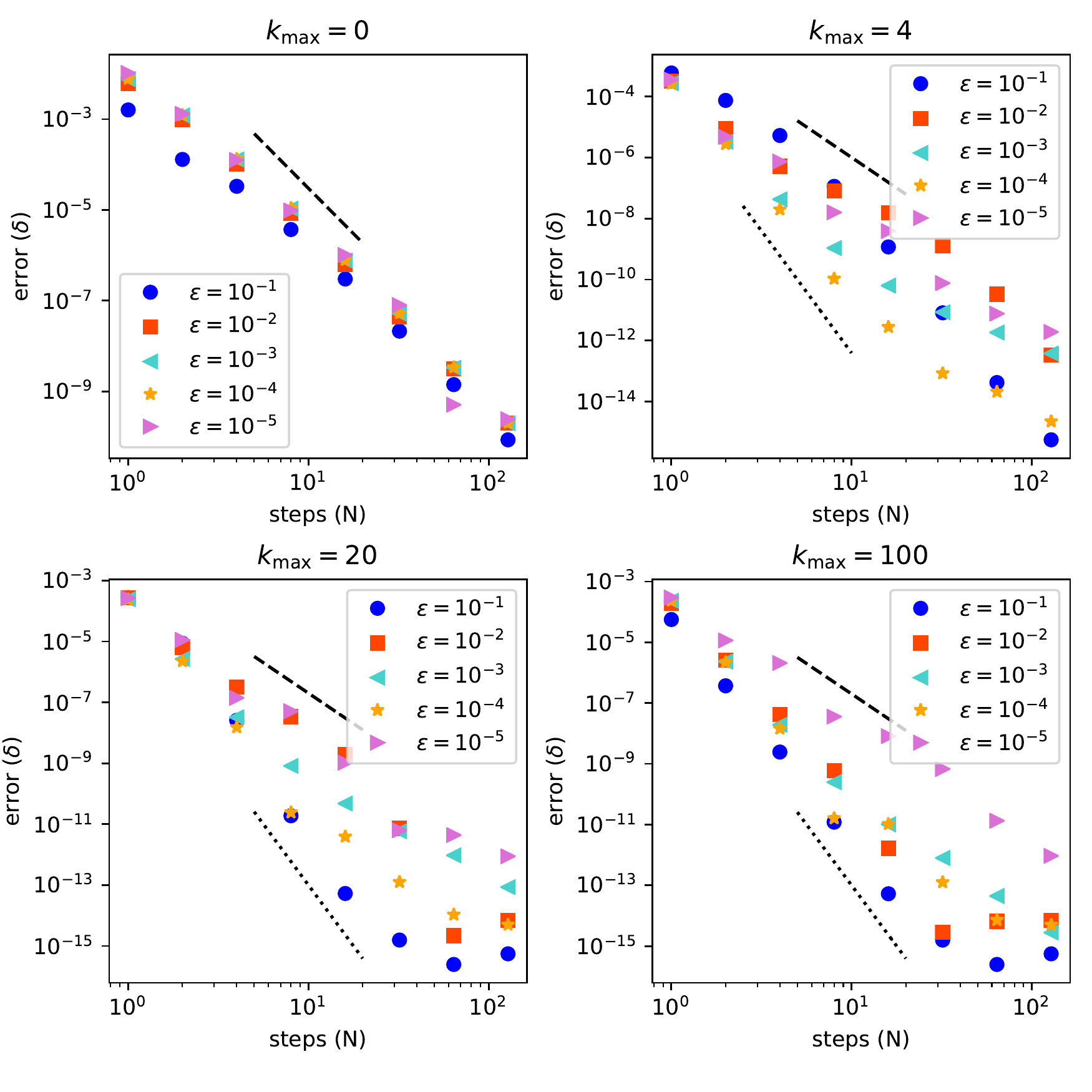}
\caption{The convergence of 8th-order Hermite IMEX schemes applied to the van der Pol equation. Dashed black lines indicate a 4th-order scaling, and dotted black lines indicate an 8th-order scaling. 
Results for $\epsilon=10^{-1}$ are plotted using blue circles, results for  $\epsilon=10^{-2}$ are plotted using vermilion squares, results for $\epsilon=10^{-3}$ are plotted using turquoise left-pointing triangles, results for $\epsilon=10^{-4}$ are plotted using orange stars, and results for $\epsilon=10^{-5}$ are plotted using purple right-facing triangles.}
\label{fig:HIMEX-8}
\end{figure}

We plot results for this test in Figures \ref{fig:HIMEX-6} and \ref{fig:HIMEX-8}. 
The schemes, when applied using the minimum $k_{\rm max}$ to achieve their nominal order, suffer from significant order reduction for some values of $\epsilon$. However, by increasing $k_{\rm max}$ to $20$ or $100$, order reduction can be fully or partially ameliorated. When comparing the efficiency of the the 6th- and 8th-order schemes to the 4th-order version, we primarily consider the cost of calculating higher-order time derivatives. Thus, we approximate that the 6th-order scheme requires about twice the total floating point operations of the 4th-order scheme, and the 8th-order scheme requires about $\sim 1.7$ times the number of floating point operations of the 6th-order scheme. 


Because the values of $k_{\rm max}$ required to reach the nominal order of the scheme differ between schemes, we compare the other choices of $k_{\rm max}$. Concerning the predictor alone, $k_{\rm max}=0$, the 4th-order algorithm presented in \cite{2020arXiv200108268S} reaches an error of $\sim10^{-10}$ using about 17000 time steps. The 6th-order algorithm reaches similar error using about $500$ time steps, while the 8th-order algorithm reaches similar errors using about $150$ time steps. Thus, in this scenario, the 6th-order scheme is roughly $\sim17$ times more efficient than the 4th-order scheme, and the 8th-order scheme is roughly $\sim33$ times more efficient than the 4th-order scheme. Naturally, the higher-order schemes would be even more beneficial if small enough time steps were used to achieve rounding-dominated errors. 

Concerning $k_{\rm max}=20$, the 4th-order scheme experiences little order reduction and reachs rounding-limited errors using around 1000 steps. The 6th-order schemes suffer order reduction for some values of $\epsilon$, but still achieve orders better than the predictor alone, reaching rounding-limited errors using $\sim 70--500$ steps depending on the value of $\epsilon$. Thus, 6th-order scheme ranges from roughly $\sim 7$ time more efficient to having about the same overall cost as the 4th-order scheme. The 8th-order scheme behaves similarly, although it suffers more extreme order reduction for some values of $\epsilon$. For some values of $\epsilon$, the 8th-order scheme reaches errors around $\sim10^{-15}$ using just 32 time steps, and is more efficient than the 6th- or 4th-order schemes, while for other values of $\epsilon$ the 8th-order scheme is less efficient. Using $k_{\rm max}=100$, the 6th-order scheme suffers virtually no order reduction and is more efficient than the 4th-order scheme. The 8th-order scheme still suffers order reduction for many values of $\epsilon$ even when using $k_{\rm max}=100$, and is generally about as efficient as the 6th-order scheme in this case, although this depends on $\epsilon$.

\subsection{Linear Initial Value Problem}\label{Linear}
Although we observed order reduction down to the order of the predictor while applying the 8th-order Hermite IMEX method for both $k_{\rm max}=20$ and $k_{\rm max}=100$ for various values of $\epsilon$ in the van der Pol oscillator problem, we show that this does not preclude the application of higher-order methods to all problems. We demonstrate this by examining a stiff linear system, finding that the 10th- and 12th- order schemes performed at or around their nominal order accuracy for $k_{\rm max}=20$.

\begin{figure}[h]
\includegraphics[width=\columnwidth]{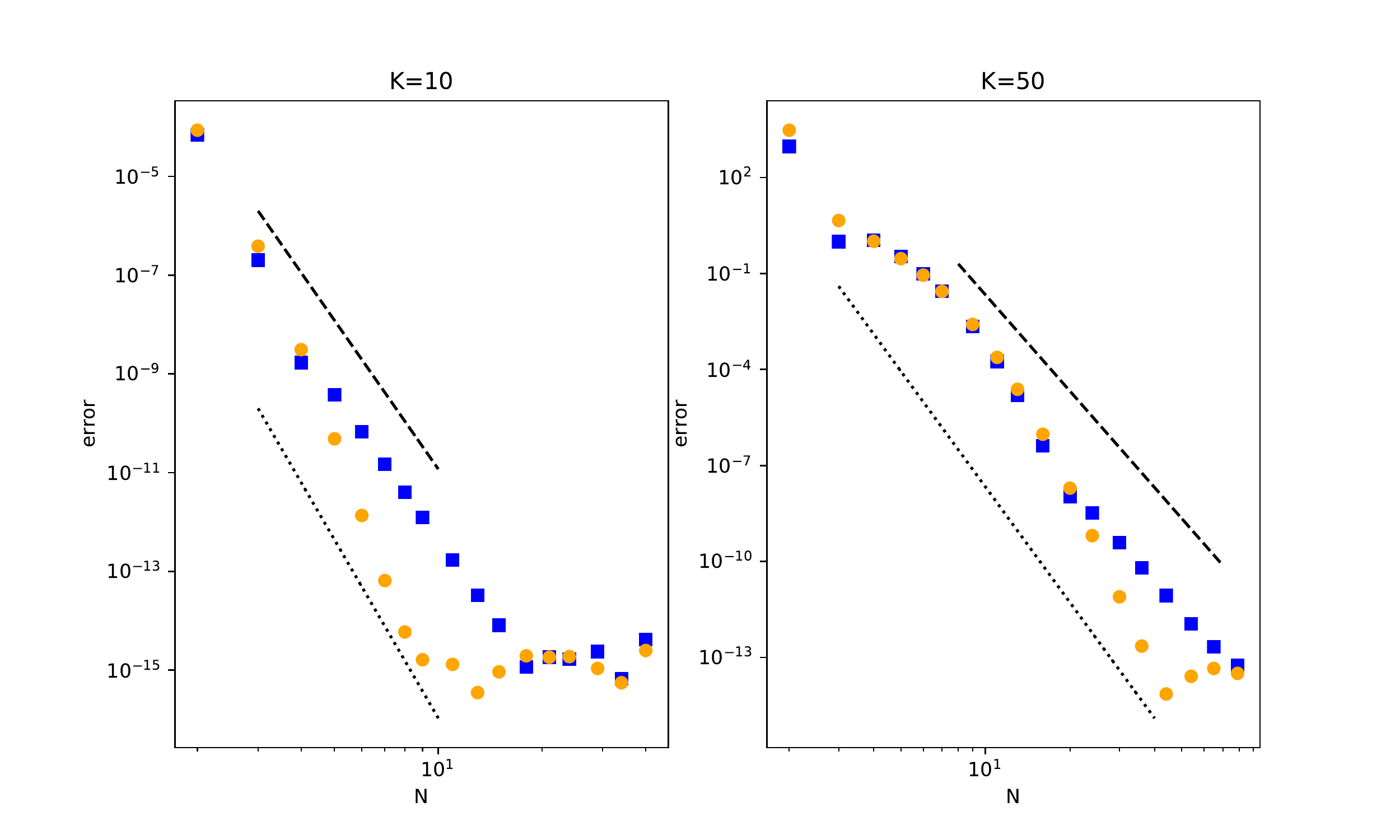}
\caption{Convergence of the 10th- and 12th-order versions of Algorithm \ref{alg 2} applied to Equation (\ref{eq:linear}), using $k_{\rm max}=20$. Results for the 12th-order algorithm are plotted using orange circles, and results for the 10th-order algorithm are plotted using blue squares.}
\label{fig:linear}
\end{figure}

For this test, we numerically solve the equation 
\begin{equation}\label{eq:linear}
\frac{dy}{dt}=-Ky,
\end{equation}
with initial value $y(0)=1$,
which is trivial to solve analytically but can be difficult to solve numerically for large values of $K$. We solve for the value of $y$ at $t=0.5$, and compare to the true value of $\exp(-K/2)$ by calculating the fractional error. 
In Figure \ref{fig:linear}
we present the results for the 10th- and 12th-order algorithms using $k_{\rm max}=20$ in all cases. We follow the methodology presented in in Algorithm \ref{alg 2}, neglecting the explicit components.
We find that although there is some scatter, the order of convergence tracks the nominal value until rounding errors begin to dominate. Notably, for this problem the cost of evaluating higher-order derivatives is trivial. For both values of $K$ tested, the 12th-order algorithm is preferable for reaching small errors, but for lenient tolerances the 10th-order algorithm may be preferable. 

\subsection{Burgers' equation}\label{Burgers}
In this section, we test the efficacy of the 6th-order Hermite IMEX scheme for solving nonlinear PDEs. We solve Burgers' equation, 
\begin{equation}\label{burgers}
\frac{\partial u}{\partial t} = -u\frac{\partial u}{\partial x} + \nu\frac{\partial^2 u}{\partial x^2}.
\end{equation}
This equation is in the form of Algorithm \ref{alg 2}, where the second term on the right-hand side of Equation (\ref{burgers}) can lead to stringent time-step constraints. Specifically, for an explicit scheme, the convective term typically requires a time step $\Delta t \lesssim \Delta x /u$, while the parabolic term typically requires $\Delta t \lesssim \Delta x^2/\nu$.

We solve Equation (\ref{burgers}) according to Algorithm \ref{alg 2} subject to periodic boundary conditions, taking derivatives using a Fourier pseudospectral method. We use an exponential filter for de-aliasing \cite{394f75eb7ba74a529384b1b8ddfedfbf}, and discretize the domain into $N_x = 64$ and $N_x = 256$ nodes. 
It is useful to investigate different spatial resolutions because evolving the diffusive term becomes more challenging compared to the convective term as resolution increases.
We calculate higher time derivatives of Equation (\ref{burgers}) using the Cauchy-Kowalevski, or Lax-Wendroff, procedure, converting time derivatives to spatial derivatives but limiting ourselves to problems without discontinuities in $u$ or its derivatives. To complete the picture, 
we specify initial conditions as 
\begin{equation}
u(x) = 2.0 + \frac{1}{4}\sin(2\pi x/L)
\end{equation}
where $L$ is the length of the domain, which we set to $1$. We evolve the system until $t = 0.15$ using various values of $\nu$. We quantify the time step in terms of $\alpha\equiv\Delta t\max(u)/\Delta x$, where $\Delta x \equiv L/N_x$ and $\alpha$ determines the step size.

\begin{figure}[h!]
\includegraphics[width=\columnwidth]{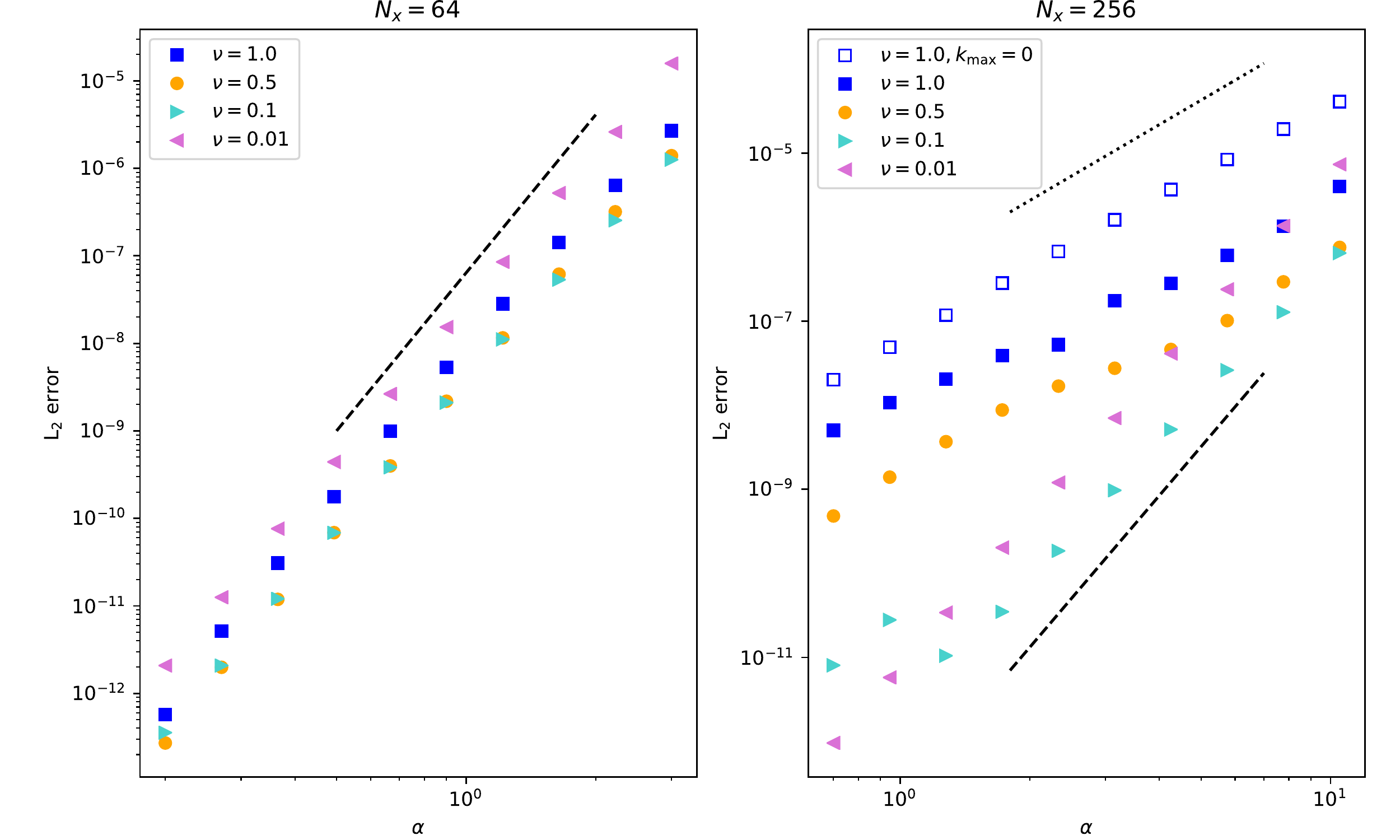}
\caption{$L_2$ errors as a function of time step when solving Burgers' equation using a 6th-order Hermite IMEX scheme. The left panel plots results computed using a resolution of $N_x=64$, while the right panel plots results using a spatial resolution of $N_x=256$. In both panels, blue squares plot results for $\nu=1$, orange circles plot results for $\nu=0.5$, turquoise right-facing triangles plot results for $\nu=0.1$, and purple left-facing triangles plot results for $\nu=0.01$. Solid markers indicate results that were computed using $k_{\rm max}=3$, while open symbols indicate results computed using $k_{\rm max}=0$. Dotted black lines indicate a 3rd-order scaling, while dashed black lines indicate a 6th-order scaling.}
\label{fig:burgers}
\end{figure}

We calculate errors by running an additional simulation as a reference, using a time step half that of the smallest used in the presented results. We then calculate the $L_2$ norm of the difference between each solution and the reference solution, plotting the results in Figure \ref{fig:burgers}. 
Concerning the $N_x=64$ tests, ${\Delta x}/{u} \sim {1}/{128}$, while the minimum ${\Delta x^2}/{\nu}$ was ${1}/{4096}$, about $32$ times smaller, and the method converged at 6th order using $k_{\rm max}=3$ in all cases. Concerning the $N_x=256$ tests, ${\Delta x}/{u} \sim {1}/{512}$, while the minimum ${\Delta x^2}/{\nu}$ was ${1}/{65536}$, about $\sim 128$ times smaller. For more extreme time scale differences, the Hermite IMEX method converges at third order using $k_{\rm max}=3$, although for smaller values of $\nu$ the method still converges at 6th order. Even when suffering order reduction, applying a few corrector iterations still reduces the error by more than an order of magnitude. 

\section{Conclusions}
In this work, we have presented a class of multiderivative IMEX schemes of arbitrary order, focusing on orders 6--12 following recent work on the 4th-order method of this family \cite{2020arXiv200108268S}. These schemes have a number of advantageous qualities. Although calculating higher-order derivatives can be expensive, such calculations are highly parallelizable, and may be advantageous on current and future parallel architectures. Additionally, many high-order multistep and Runge-Kutta methods require storing a number of intermediate stages of the solution, while the methods presented here only require storing values at the beginning and end of a time step, regardless of the order of the algorithm. However, because higher-order schemes appear more susceptible to order reduction for more stiff problems, lower-order schemes can be more efficient depending on the problem at hand. Similarly, tuning $k_{\rm max}$ adaptively may be fruitful. We note that the 4th-order version of these schemes has been proven to be asymptotic preserving \cite{2020arXiv200108268S}, although we have not extended this proof to arbitrary orders in this work.

For nonlinear problems, especially systems of nonlinear PDEs, computing higher-order time derivatives can be very expensive. This is one reason that the original manner of implementing ADER schemes, based on the Cauchy-Kowalevski procedure and forward Taylor approximations  \cite{TT02}, has fallen out of favour compared to implementations that implicitly solve a weak formulation of the PDE in time \cite{2008JCoPh.227.3971D}. However, the methods presented here can handle stiff terms implicitly, and also require significantly fewer time derivatives of the governing PDE to be calculated to achieve an algorithm of a given order. Thus, the algorithms presented here have a number of benefits, and do not suffer the same drawbacks as previous very-high-order schemes.

\section*{Acknowledgements}
The author is thankful to Cole Miller and Derek Richardson for their feedback during the preparation of this manuscript, and to Bill Dorland and Keigo Nitadori for their comments, all of which improved the quality of this work.
We gratefully acknowledge support from NASA under grant NNX17AF29G.

\pagebreak
\appendix

\section{Hermite schemes up to 12th order}\label{apndx:A}
6th:
\begin{equation}
\Delta u = \frac{\Delta t}{2}(f_1+f_0)-\frac{\Delta t^2}{10}(f^{(1)}_1-f^{(1)}_0)+\frac{\Delta t^3}{120}(f^{(2)}_1+f^{(2)}_0)
\end{equation}

8th:
\begin{equation}
\begin{split}
\Delta u = 
\frac{\Delta t}{2}(f_1+f_0)
-\frac{3\Delta t^2}{28}(f^{(1)}_1-f^{(1)}_0)
+\frac{\Delta t^3}{84}(f^{(2)}_1+f^{(2)}_0)\\
-\frac{\Delta t^4}{1680}(f^{(3)}_1-f^{(3)}_0)
\end{split}
\end{equation}

10th:
\begin{equation}
\begin{split}
\Delta u = 
\frac{\Delta t}{2}(f_1+f_0)
-\frac{\Delta t^2}{9}(f^{(1)}_1-f^{(1)}_0)
+\frac{\Delta t^3}{72}(f^{(2)}_1+f^{(2)}_0)\\
-\frac{\Delta t^4}{1008}(f^{(3)}_1-f^{(3)}_0)
+\frac{\Delta t^5}{30240}(f^{(4)}_1+f^{(4)}_0)
\end{split}
\end{equation}

12th:
\begin{equation}
\begin{split}
\Delta u = 
\frac{\Delta t}{2}(f_1+f_0)
-\frac{5\Delta t^2}{44}(f^{(1)}_1-f^{(1)}_0)
+\frac{\Delta t^3}{66}(f^{(2)}_1+f^{(2)}_0)\\
-\frac{\Delta t^4}{792}(f^{(3)}_1-f^{(3)}_0)
+\frac{\Delta t^5}{15840}(f^{(4)}_1+f^{(4)}_0)
-\frac{\Delta t^6}{665280}(f^{(5)}_1-f^{(5)}_0)
\end{split}
\end{equation}

\bibliography{mybibfile}

\end{document}